\documentclass[conference]{IEEEtran}
\IEEEoverridecommandlockouts
\usepackage{cite}
\usepackage{amsmath,amssymb,amsfonts}
\usepackage{algorithmic}
\usepackage{graphicx}
\usepackage{textcomp}
\usepackage{xcolor}
\usepackage{booktabs, multirow,array}
\usepackage{float}
\usepackage{subfig}

\def\BibTeX{{\rm B\kern-.05em{\sc i\kern-.025em b}\kern-.08em
    T\kern-.1667em\lower.7ex\hbox{E}\kern-.125emX}}
\begin{document}

\title{Map-Assisted Material Identification at 100 GHz and Above Using Radio Access Technology
}
\makeatletter
\newcommand{\linebreakand}{%
  \end{@IEEEauthorhalign}
  \hfill\mbox{}\par
  \mbox{}\hfill\begin{@IEEEauthorhalign}
}
\makeatother
\author{\IEEEauthorblockN{Yi Geng}
	\IEEEauthorblockA{\textit{UNISOC Research, China} \\
		yi.geng@unisoc.com}
}
\maketitle

\begin{abstract}
The inclusion of material identification in wireless communication system is an emerging area that offers many opportunities for 6G systems. By using reflected radio wave to determine the material of reflecting surface, not only the performance of 6G networks can be improved, but also some exciting applications can be developed. In this paper, we recap a few prior methods for material identification, then analyze the impact of thickness of reflecting surface on reflection coefficient and present a new concept ``settling thickness'', which indicates the minimum thickness of reflecting surface to induce steady reflection coefficient. Finally, we propose a novel material identification method based on ray-tracing and 3D-map. Compared to some prior methods that can be implemented in single-bounce-reflection scenario only, we extend the capability of the method to multiple-bounce-reflection scenarios.
\end{abstract}


\section{Introduction}

Using radio access technology to identify the material of scatterers has the potential to become an essential component in future 6G networks. By predicting the materials of scatterers, 6G system will be able to understand the users' environment and take actions to improve the performance of the network\cite{b1}\cite{b2}. For example, material information of scatterers can be used to predict propagation loss. If a scatterer is identified as metal surface that has very low reflection loss (RL), proactive decision about transmit power to the metal surface can be made for interference mitigation purpose. 3D-map with material information can find applications in digital twin and autonomous driving. For example, by detecting ice on the road, autonomous vehicle can change its driving operation to a more safe mode. However, most of the data in conventional 3D-map are the culmination of the objects as well as color information, but without any material information, since the existing 3D scanning technologies such as simultaneous-localization-and-mapping are not capable of collecting material information. On the other hand, material information in an environment may be time-variable. In other words, material information of the objects may change over time and lead to out-of-date material information in 3D-map. For example, the material information in a digitized road map collected during good weather may not always be accurate enough to portray the real road condition all the time, e.g., snowy or wet road in bad weather. Therefore, extremely reliable real-time material identification method is crucial to support some sensing use cases, e.g., autonomous driving.

\begin{figure}[htb]
	\centerline{\includegraphics[width=0.9\linewidth, height=10cm, keepaspectratio]{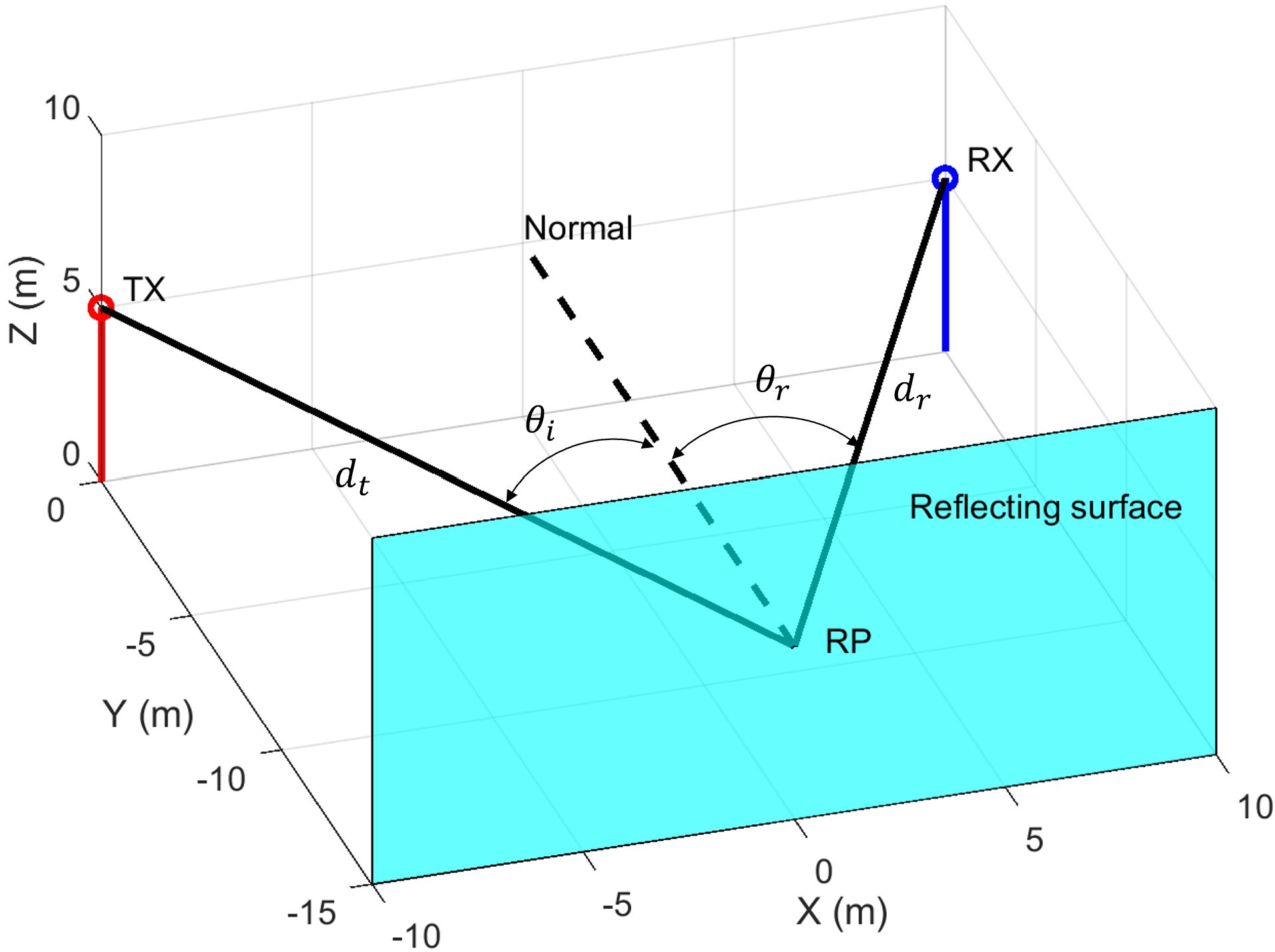}}
	\caption{An illustration of how RL is calculated.}
	\label{fig_1}
\end{figure}

\begin{figure}[t]
	\centerline{\includegraphics[width=0.9\linewidth, height=10cm, keepaspectratio]{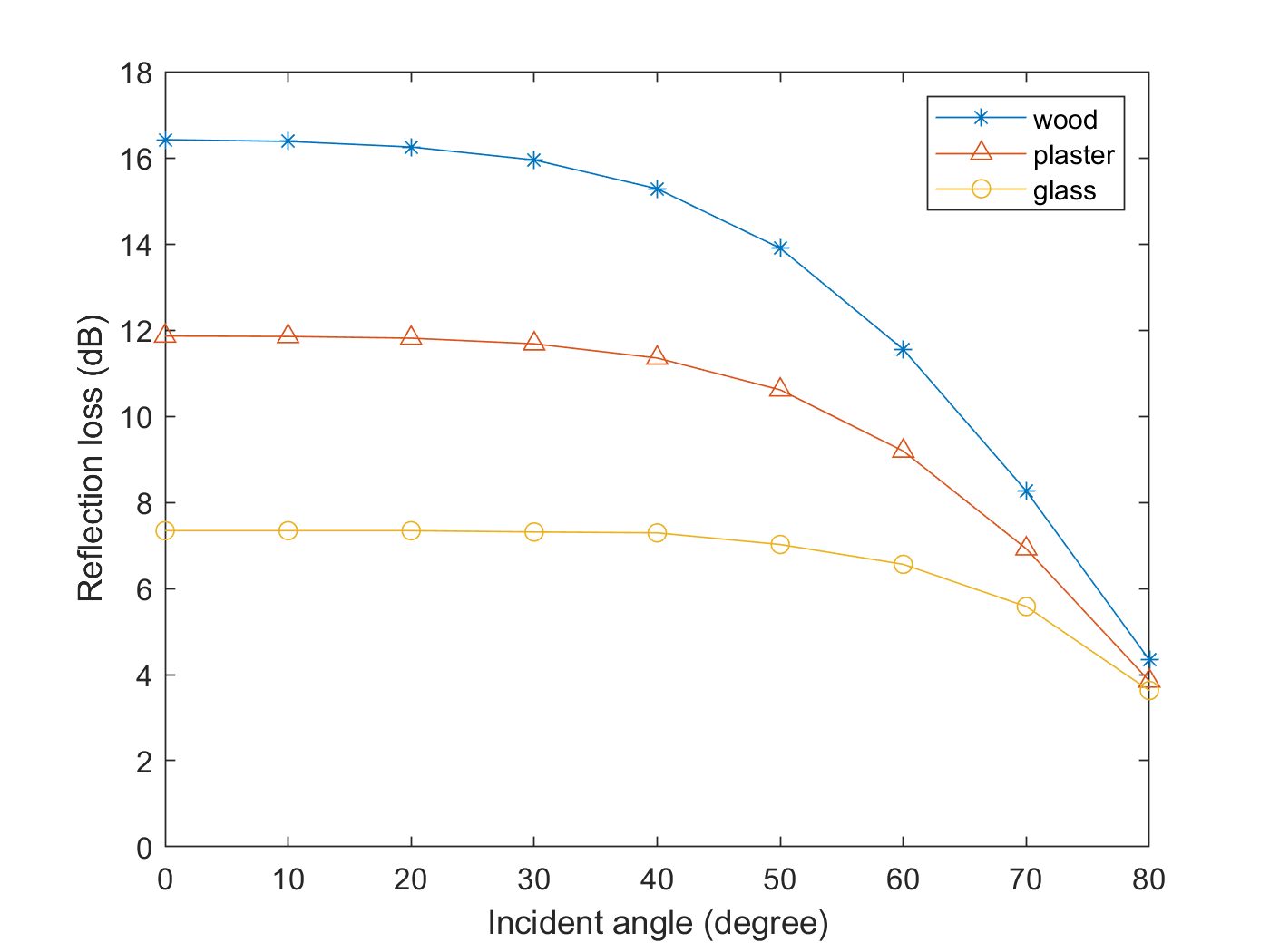}}
	\caption{RL induced by reflecting surfaces made of wood, plaster, and glass at 100 GHz.}
	\label{fig_2}
\end{figure}

\section{Related Work}
\begin{table*}[htb]
	\centering
	\caption{A database of RL induced by reflecting surfaces made of wood, plaster, and glass at 100 GHz}\label{tab_2}
	\begin{tabular}{@{} l*{9}{>{$}c<{$}} @{}}
		\toprule
		\textbf{Material} & \multicolumn{9}{c@{}}{\textbf{RL in decibel at different incident angle in degree}}\\
		\cmidrule(l){2-9}
		& \mathrm{0\textdegree} &  \mathrm{10\textdegree} &  \mathrm{20\textdegree} &  \mathrm{30\textdegree} &  \mathrm{40\textdegree} &  \mathrm{50\textdegree} &  \mathrm{60\textdegree} &  \mathrm{70\textdegree} &  \mathrm{80\textdegree} \\
		\midrule
		Wood & 16.42 & 16.38 & 16.25 & 15.95 & 15.28 & 13.9 & 11.55 & 8.26 & 4.34 \\[1ex]	
		
		Plaster & 11.86 & 11.85 & 11.81 & 11.68 & 11.35 & 10.61 & 9.19 & 6.92 & 3.85 \\[1ex]
		
		Glass & 7.34 & 7.34 & 7.34 & 7.31 & 7.29 & 7.02 & 6.56 & 5.58 & 3.63 \\[1ex]
		
	\end{tabular}
\end{table*}

Most studies in the field of material identification have only focused on single-bounce-reflection\cite{b3}\cite{b4}, that is, it is assumed that single-bounce-reflection trajectories are always available (single-bounce-assumption) \cite{b5}. For example, in \cite{b3}, the authors demonstrated how permittivity obtained from reflected radio signals can be used to identify the material of reflecting surface. The condition that applies the method in \cite{b3} properly is the existence of single-bounce-reflection trajectories. However, single-bounce-reflection trajectory may not always exist. Therefore, single-bounce-assumption-based approaches are not applicable to sensing systems that are deployed in rich scattering environments, since the cluttered objects in such environments induce a large number of multiple-bounce-reflection paths but few single-bounce-reflection paths. The radio waves along single-bounce-reflection trajectories may not be sufficient to characterize the properties of reflecting surfaces for sensing purpose.

My previous study in \cite{b4} has analyzed the impact of material, incident angle, and frequency on RL, and proposed a method to obtain RL: As the simulation illustrated in Fig.~1, a radio wave transmitted by the TX strikes a reflecting surface at a reflection point (RP) with incident angle $\theta_i$. The received power at the RX, which is attenuated by free space propagation and reflection, is measured by the RX. The overall path loss (PL) including free space path loss (FSPL) and RL induced by the reflecting surface can be calculated by transmit power $P_{TX}$ at TX minus received power $P_{RX}$ at RX:

\begin{equation}\label{eqn_1}
	PL = P_{TX} - P_{RX}.
\end{equation}

The FSPL over length of trajectory $d$ ($d_t+d_r$) can be calculated by Friis equation:

\begin{equation}\label{eqn_2}
	FSPL(f,d) = 32.4 +20\log_{10}(f) + 20\log_{10}(d).
\end{equation}

Finally, the RL induced by the reflecting surface can be calculated by subtracting FSPL from PL:
 
\begin{equation}\label{eqn_3}
 	RL = PL - FSPL(f,d).
\end{equation}
 
Fig.~2 illustrates the simulated RL as a function of incident angle, for reflecting surfaces made of wood, plaster, and glass at 100~GHz. The RL data in Fig.~2 are also tabulated in TABLE~I. In the simulation, the roughness of the reflecting surfaces made of wood, plaster, and glass are 0.4~mm, 0.2~mm, and 0~mm, respectively. However, the study in \cite{b4} also focused on single-bounce-reflection.

My previous study in\cite{b6} has sought the material identification solution for double-bounce-reflection trajectories by analyzing the total RL induced by two reflecting surfaces. However the methods proposed in \cite{b6} are unable to accurately obtain the information about incident angle especially when the incident angle is small. As a consequence, high degree of accuracy cannot be achieved. The left part of the RL curves in Fig.~2 are almost flat, indicating that we cannot accurately estimate the incident angle by RL in the scenarios where small incident angles dominate the reflections, such as a self-driving car detects other vehicles ahead.

\begin{figure}[b]
	\centerline{\includegraphics[width=0.9\linewidth, height=10cm, keepaspectratio]{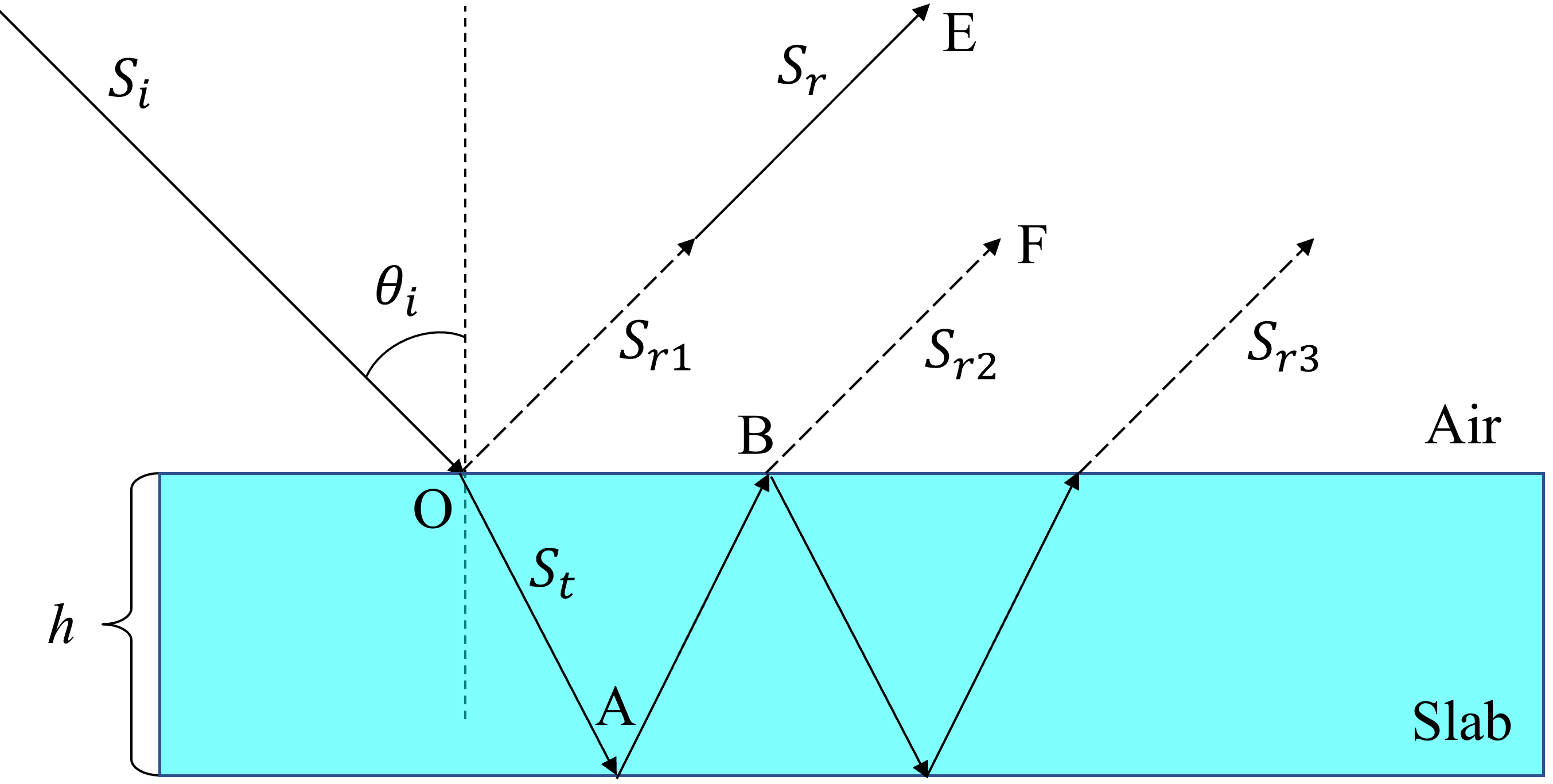}}
	\caption{Transmit and reflected radio waves for an oblique radio wave $S_i$ affected by a thin slab.}
	\label{fig_3}
\end{figure}

\section{Impact of Thickness on Reflection Coefficient}

\begin{figure*}[t]
	\centering
	\subfloat[Wood at 28~GHz]{\label{fig:h}\includegraphics[width=0.65\columnwidth]{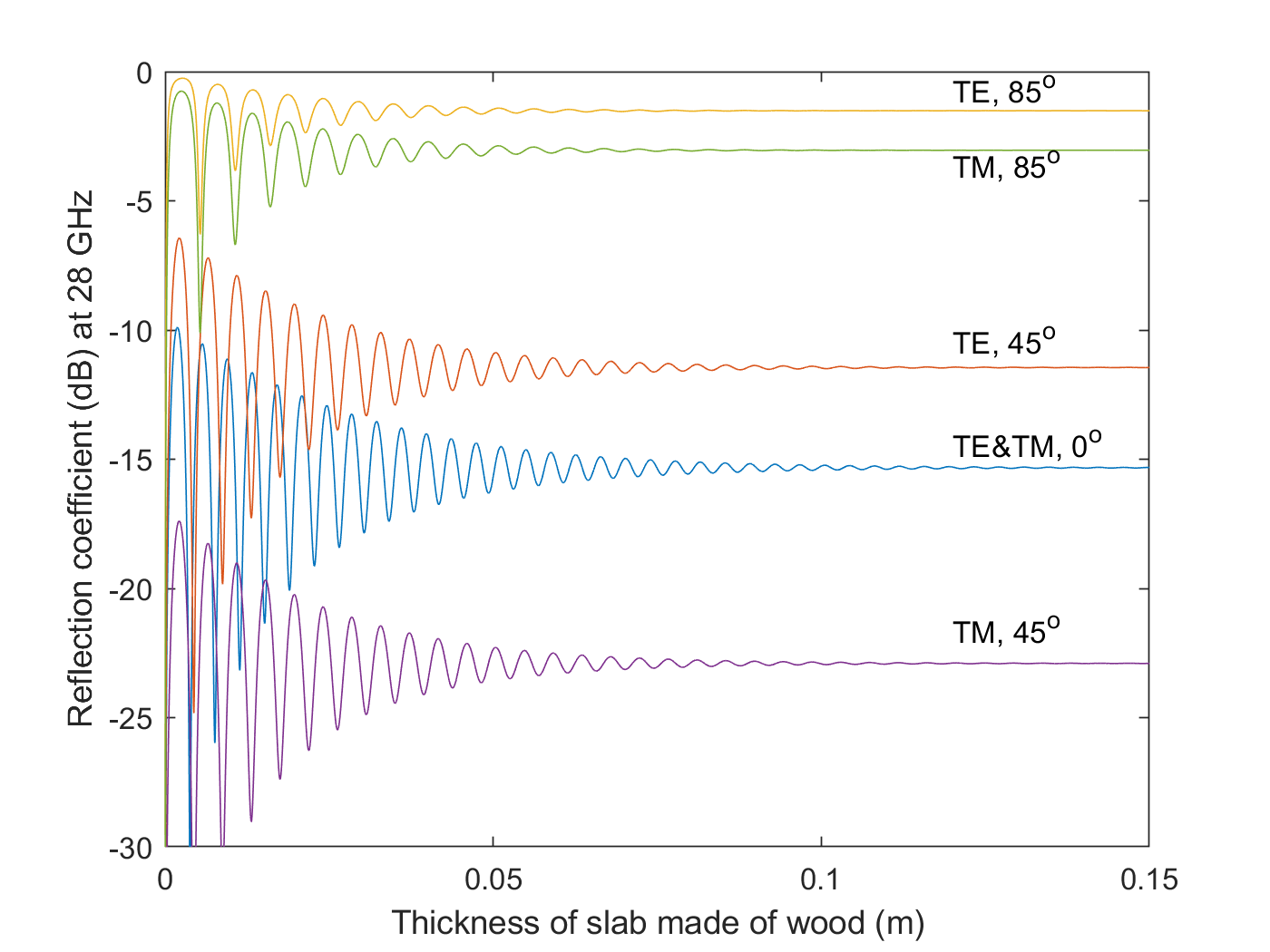}}\quad
	\subfloat[Wood at 100~GHz]{\label{fig:i}\includegraphics[width=0.65\columnwidth]{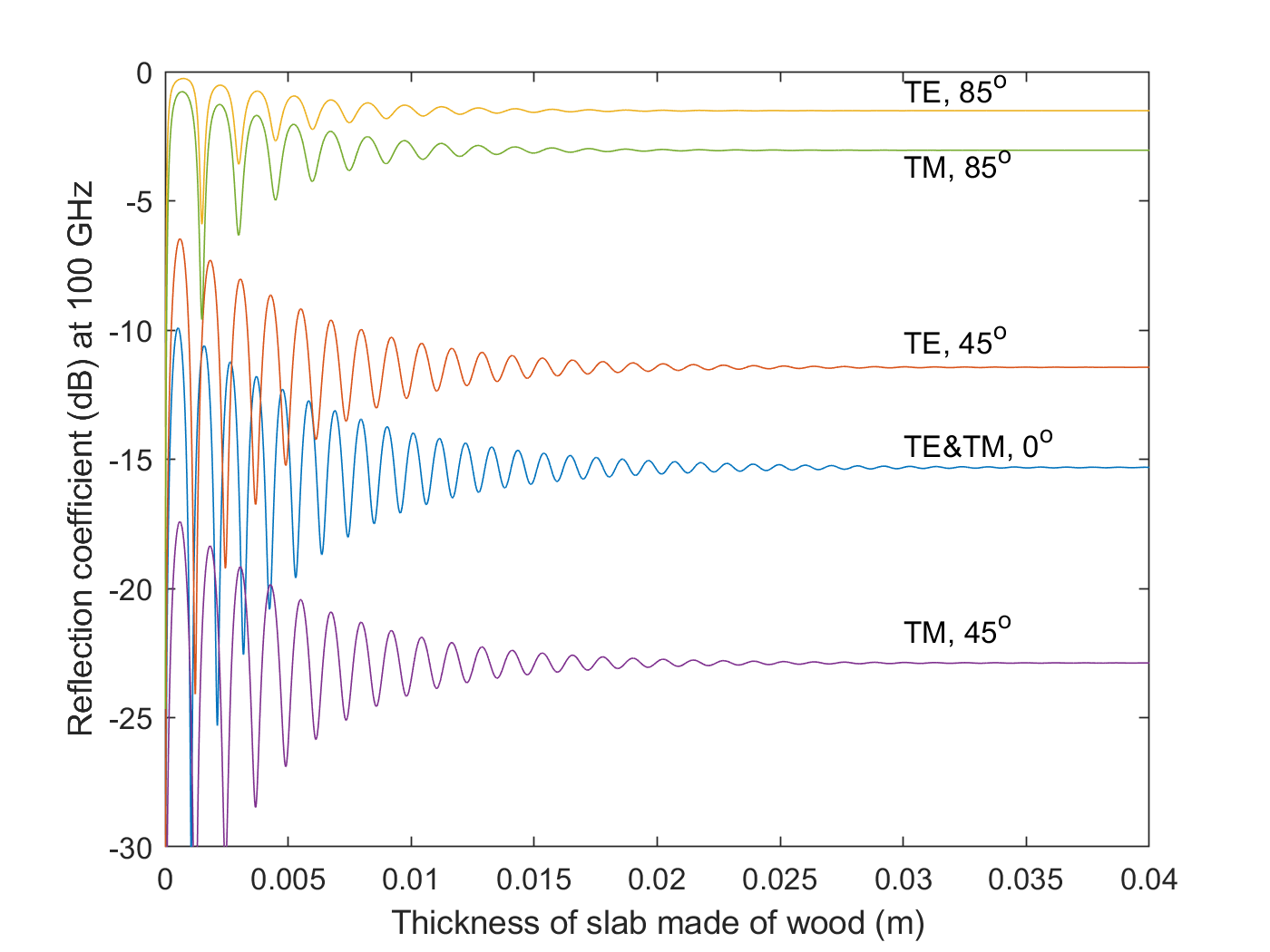}}\quad
	\subfloat[Wood at 1~THz]{\label{fig:g}\includegraphics[width=0.65\columnwidth]{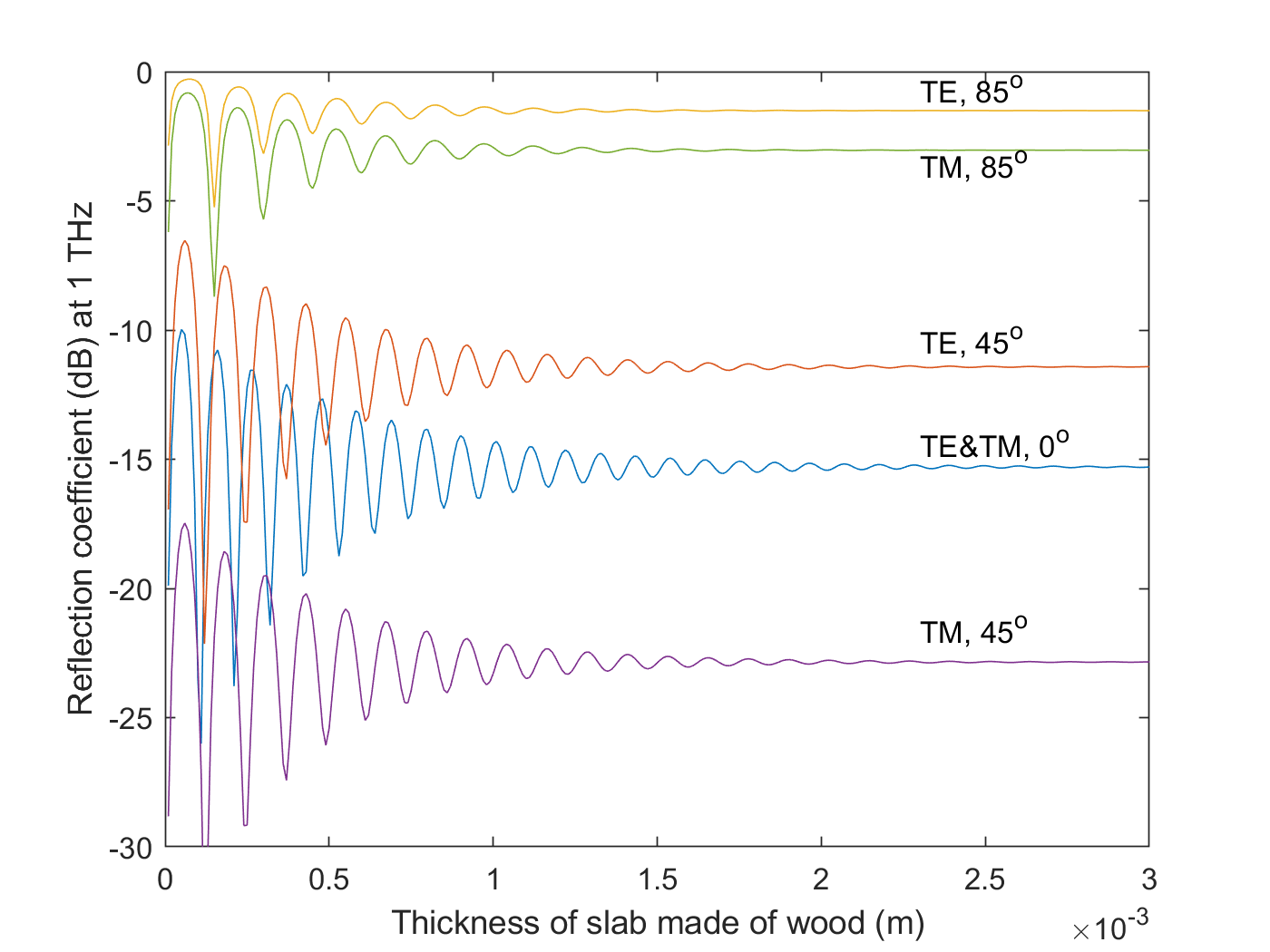}}\quad
	\subfloat[Plaster at 28~GHz]{\label{fig:h}\includegraphics[width=0.65\columnwidth]{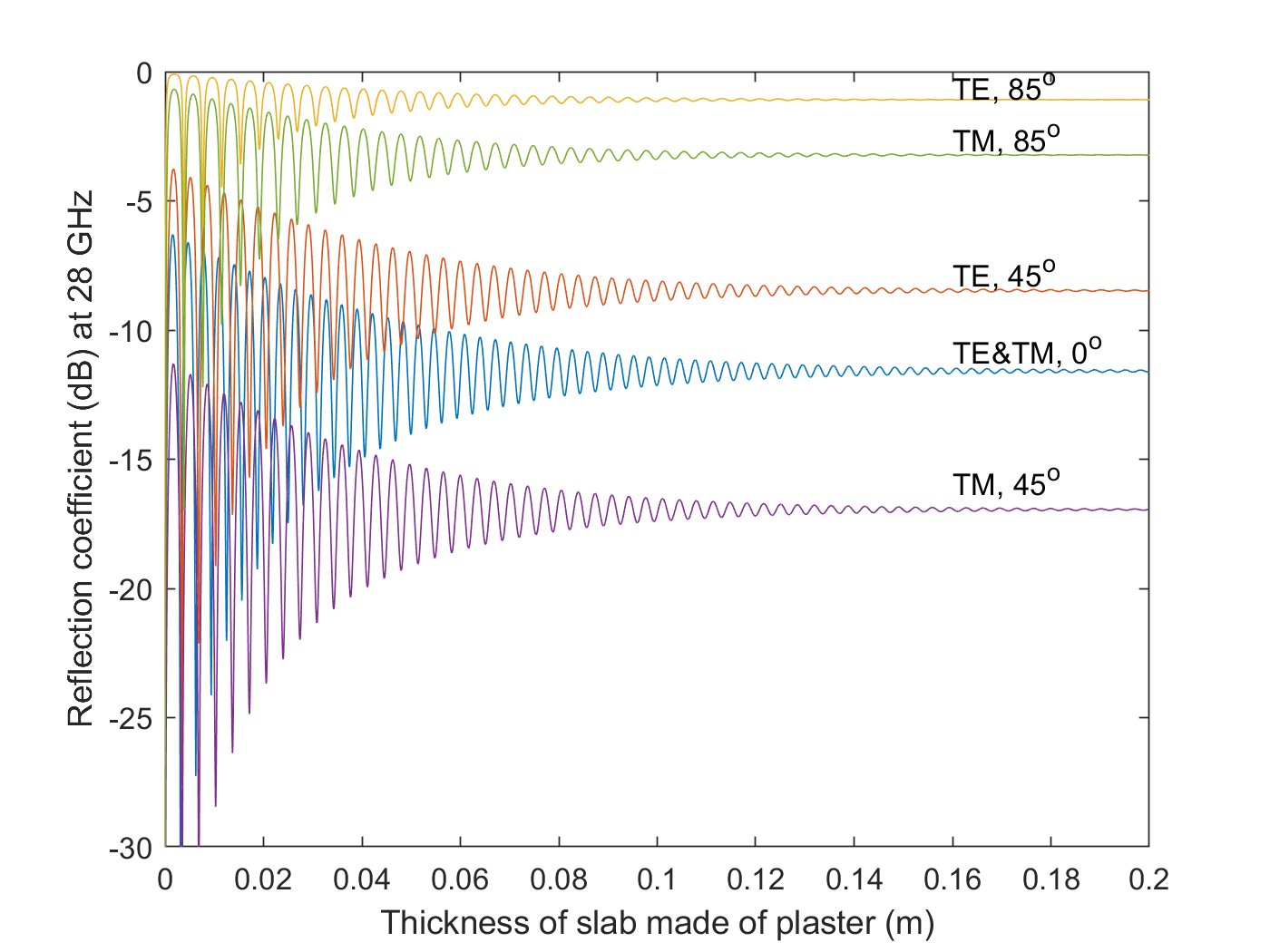}}\quad
	\subfloat[Plaster at 100~GHz]{\label{fig:i}\includegraphics[width=0.65\columnwidth]{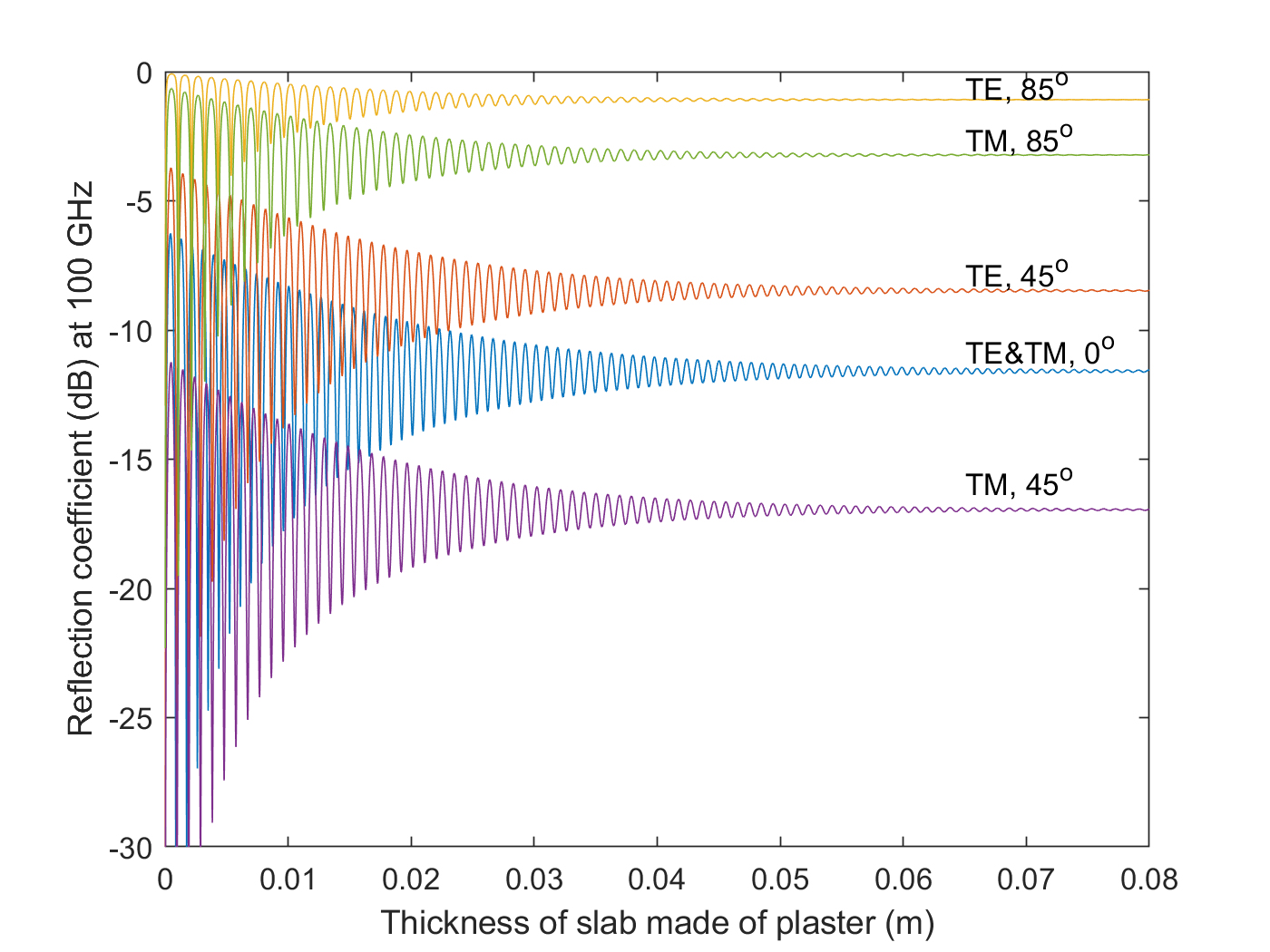}}\quad
	\subfloat[Plaster at 1~THz]{\label{fig:g}\includegraphics[width=0.65\columnwidth]{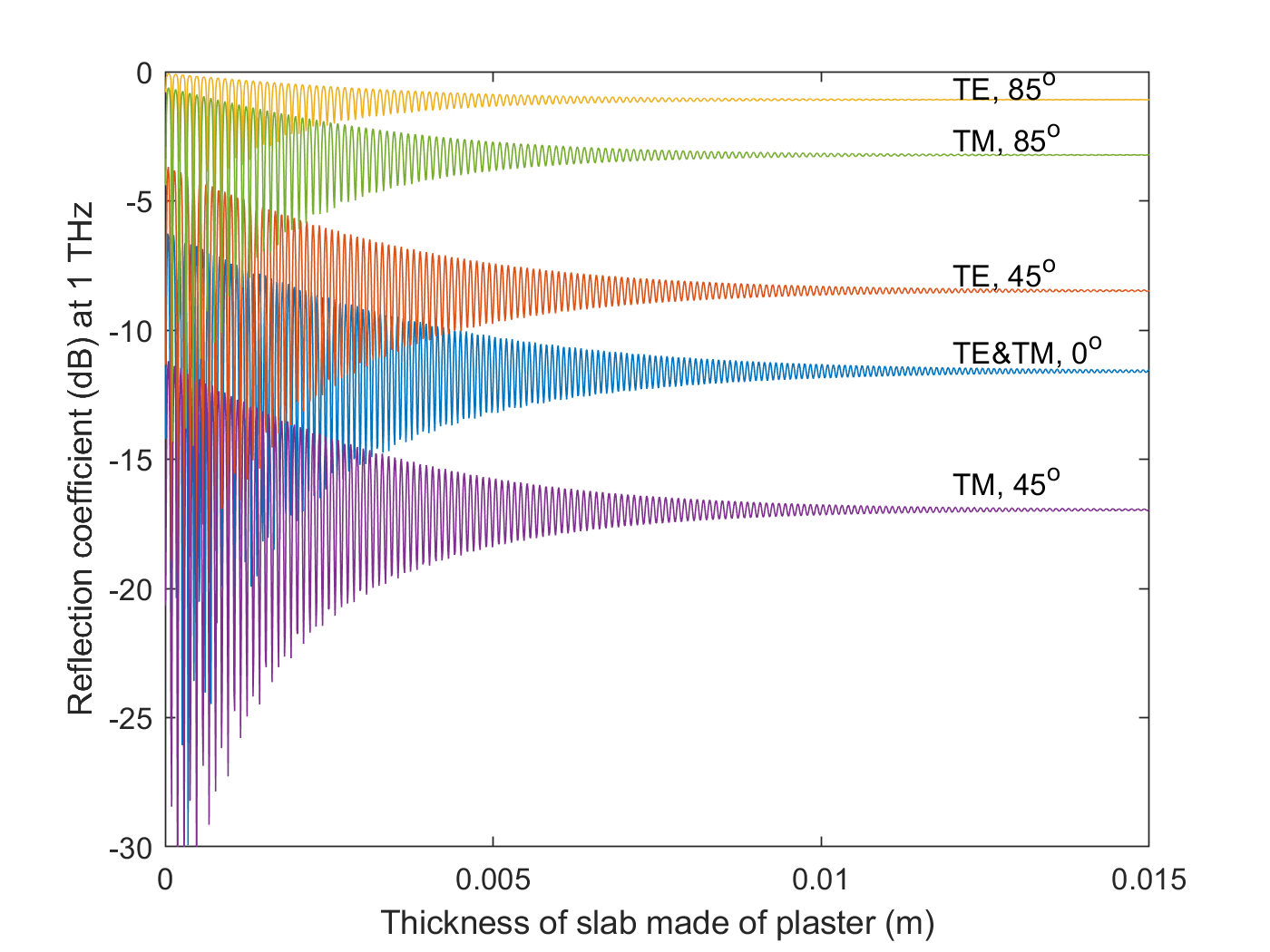}}\quad
	\subfloat[Glass at 28~GHz]{\label{fig:h}\includegraphics[width=0.65\columnwidth]{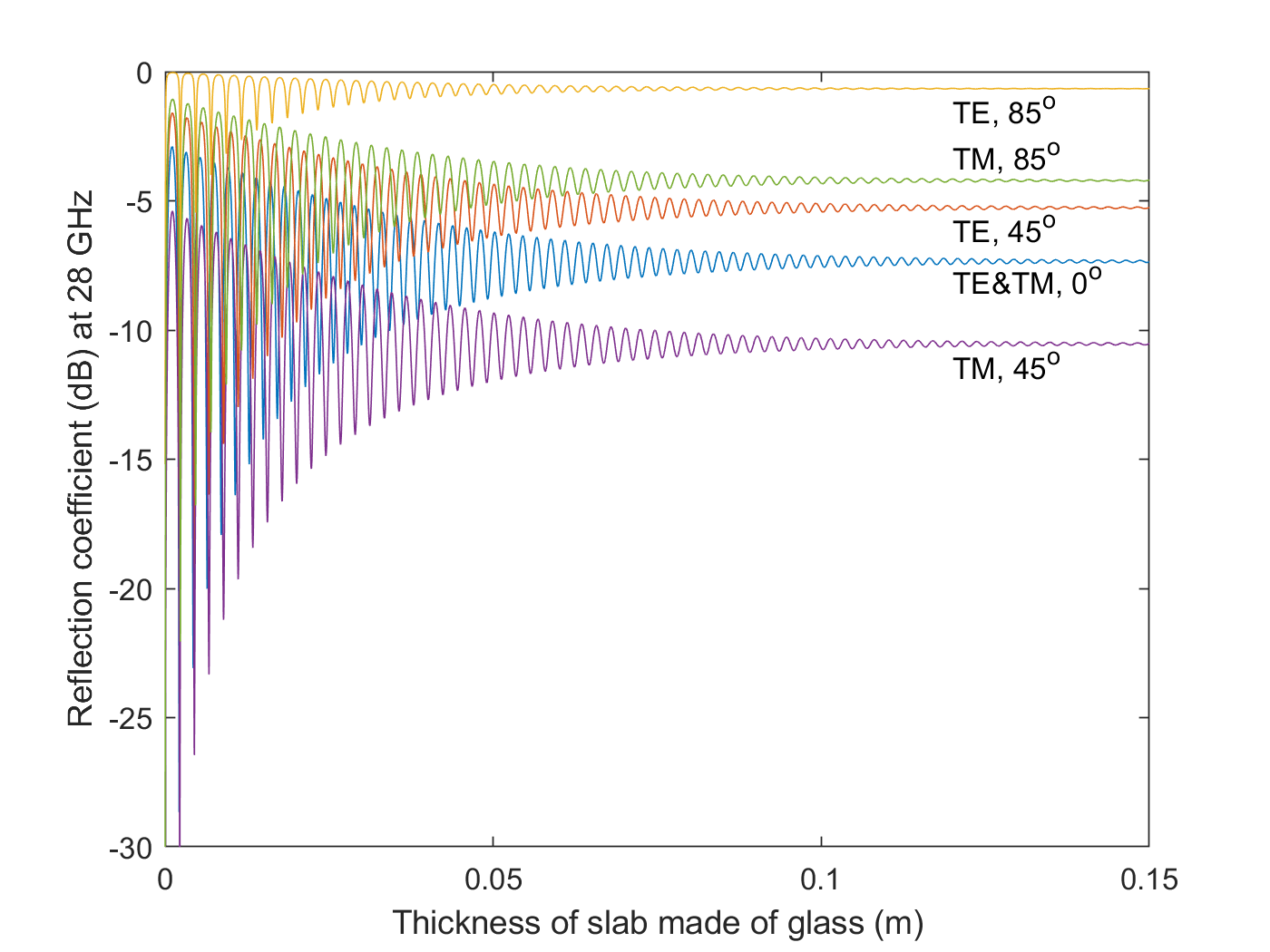}}\quad
	\subfloat[Glass at 100~GHz]{\label{fig:i}\includegraphics[width=0.65\columnwidth]{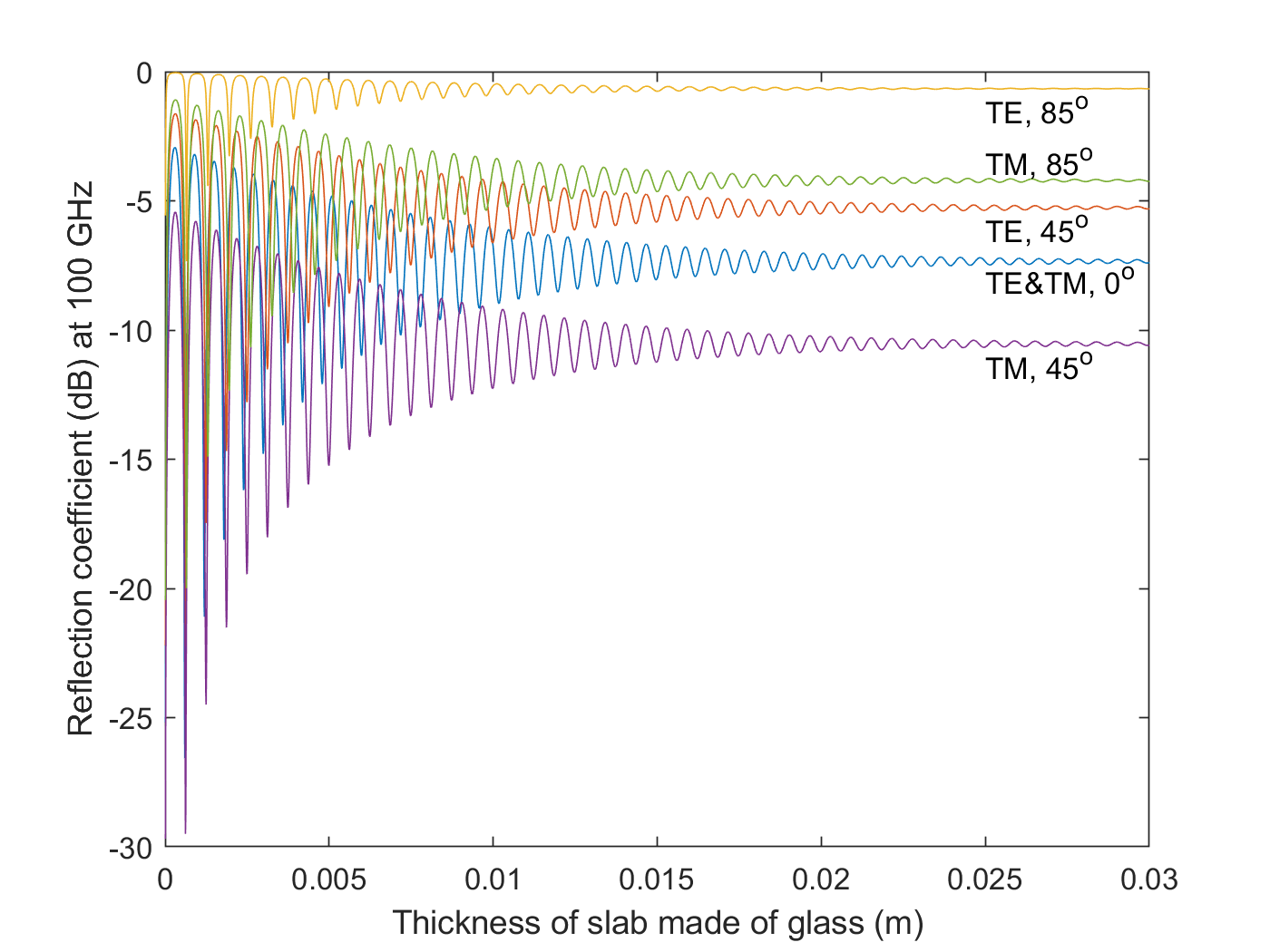}}\quad
	\subfloat[Glass at 1~THz]{\label{fig:g}\includegraphics[width=0.65\columnwidth]{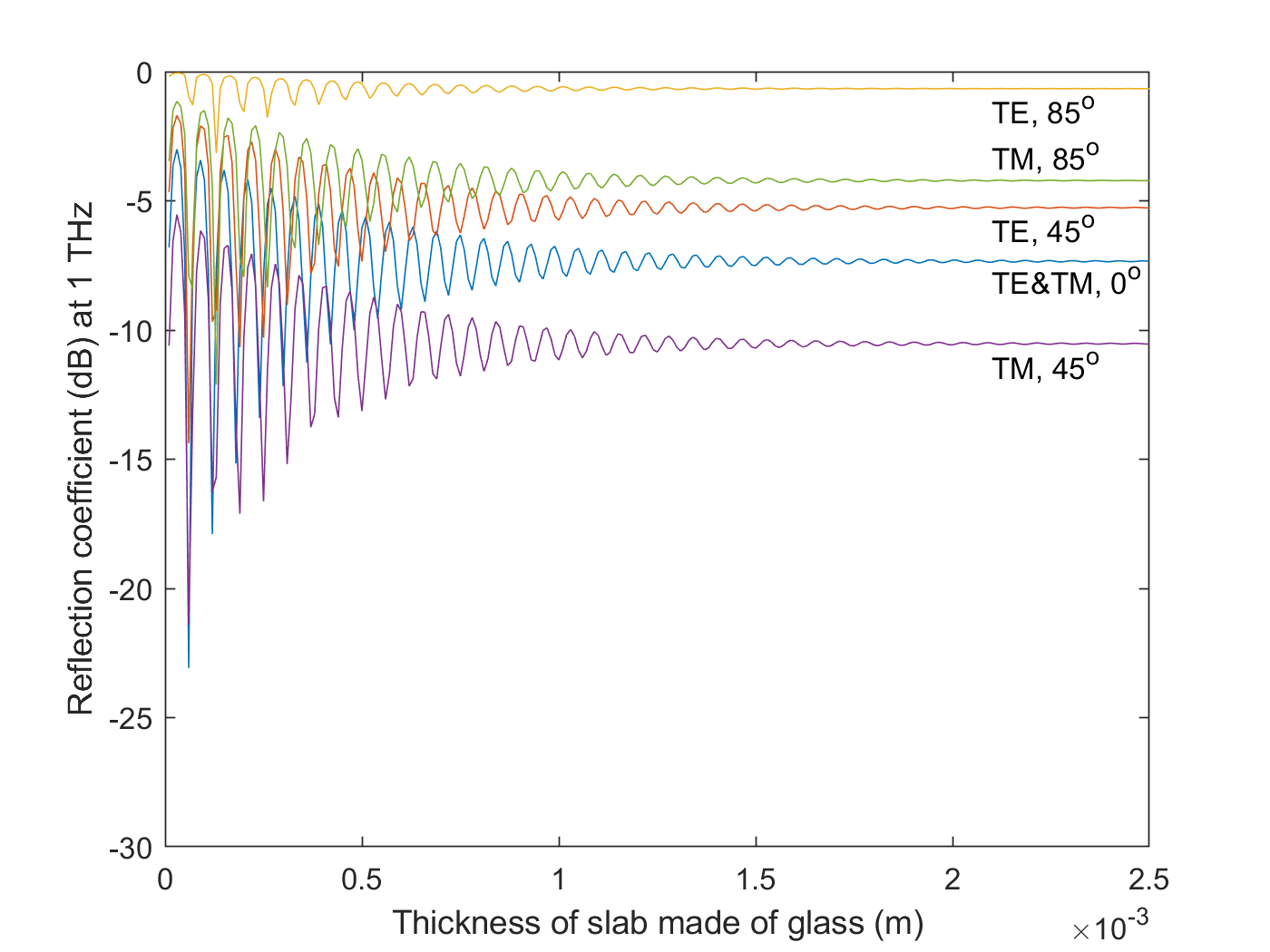}}\\	
	\caption{Reflection coefficient for TE and TM polarization for the slabs made of wood, plaster, and glass at 28~GHz, 100~GHz, and 1~THz as a function of thickness for incident angles 0\textdegree, 45\textdegree, and 85\textdegree.}
	\label{fig_4}
\end{figure*}

Reflection coefficient is defined as the ratio of amplitudes of reflected and transmit waves. This section provides theoretical analysis and simulation results of reflection coefficient for slabs with different thicknesses and different materials. As illustrated in Fig.~3, when a radio wave $S_i$ propagating in the air impinges on a thin slab with thickness $h$ at point O, it is either reflected along OE as shown by $S_{r1}$, or refracted and transmitted through the slab along OA as shown by $S_t$. At point A, the wave $S_t$ may once again be reflected along AB and then refracted along BF at point B. Multiple reflection and refraction may occur and produce a number of parallel waves $S_{r1}$, $S_{r2}$, $S_{r3}$ \dots, emerging from the top surface of the slab. $S_{r1}$, $S_{r2}$, $S_{r3}$ interfere with each other and produce a resultant wave $S_r$ along OE. $S_{r2}$ and $S_{r3}$ must travel extra distance than $S_{r1}$. Therefore, the phase shifts, which are the differences between the phase of $S_{r1}$ and the phases of $S_{r2}$ and $S_{r3}$, can be in phase, or out of phase. The degree of constructive or destructive interference between $S_{r1}$, $S_{r2}$, $S_{r3}$, which determine the amplitude of resultant wave $S_r$, depends on the difference between their phases, this difference in turn depends on the slab thickness $h$. Therefore, the thickness of reflecting surface may impact the strength of reflected wave $S_r$.

To understand how different frequencies, thicknesses, materials, and incident angles impact on reflection coefficient, we calculate the reflection coefficient for slabs made of wood, plaster, and glass. The reflection coefficient of slab for transverse electric (TE) polarization, $r'_{TE}$, and reflection coefficient of slab for transverse magnetic (TM) polarization, $r'_{TM}$, are given by:

\begin{equation}\label{eqn_4}
	r'_{TE} = \frac{r_{TE}(1-exp(-j2q))}{1-{r^2_{TE}}exp(-j2q)},
\end{equation}

\begin{equation}\label{eqn_5}
	r'_{TM} = \frac{r_{TM}(1-exp(-j2q))}{1-{r^2_{TM}}exp(-j2q)},
\end{equation}
where
\begin{equation}\label{eqn_6}
	q = 2{\pi}hf\sqrt{{\eta}-sin^2{\theta_i}},
\end{equation}
$h$ is the thickness of the slab in meter, $f$ is the carrier frequency in giga Hertz, $\theta_i$ is incident angle in radian, $\eta$ is the relative permittivity of the material of the slab and is given by \eqref{eqn_7}:
\begin{equation}\label{eqn_7}
	\eta = af^b -j17.98cf^d/f,
\end{equation}
where $a$, $b$, $c$, and $d$ are material properties that determine the relative permittivity and conductivity. $a$, $b$, $c$, and $d$ of wood, plaster, and glass recommended by ITU are tabulated in TABLE~II\cite{b7}.

\begin{table}[t]
	\caption{Parameters of three common building materials recommended by ITU-R P.2040-1}
	\begin{center}
		\begin{tabular}{ccccccc}
			\hline
			\multirow{2}{*}{\textbf{Material}} &\multicolumn{2}{c}{\textbf{Permittivity}}&\multicolumn{2}{c}{\textbf{Conductivity}} \\
			& $a$ & $b$ & $c$ & $d$\\
			\hline
			Wood &	1.99 & 0 & 0.0047 & 1.0718 \\
			Plaster & 2.94 &	0 &	0.0116 & 0.7076\\
			Glass & 6.27 & 0 & 0.0043 & 1.1925 &\\
			\hline
		\end{tabular}
	\end{center}
	\label{tab_2}
\end{table}

$r_{TE}$ and $r_{TM}$ in \eqref{eqn_4} and \eqref{eqn_5} represent Fresnel reflection coefficient for TE and TM polarization for slab sufficiently thick such that the effect of $S_{r2}$ and $S_{r3}$ is negligible. $r_{TE}$ and $r_{TM}$ are given by:

\begin{equation}\label{eqn_8}
	r_{TE} = \frac{cos{\theta_i}-\sqrt{\eta-sin^2{\theta_i}}}{cos{\theta_i}+\sqrt{\eta-sin^2{\theta_i}}},
\end{equation}

\begin{equation}\label{eqn_9}
	r_{TM} = \frac{{\eta}cos{\theta_i}-\sqrt{\eta-sin^2{\theta_i}}}{{\eta}cos{\theta_i}+\sqrt{\eta-sin^2{\theta_i}}}.
\end{equation}

Fig.~4 illustrates the reflection coefficient $r'_{TE}$ and $r'_{TM}$ calculated by \eqref{eqn_4}-\eqref{eqn_9} at 28~GHz, 100~GHz, and 1~THz as a function of thickness of the slab for incident angles 0\textdegree, 45\textdegree, and 85\textdegree. We can see that the curves of $r'_{TE}$ and $r'_{TM}$ fluctuate less and less and finally converge to $r_{TE}$ and $r_{TM}$ as the thickness of the slab increases, because if we substitute $h=+\infty$ into \eqref{eqn_4}-\eqref{eqn_6}, as a result we get $r'_{TE}=r_{TE}$ and $r'_{TM}=r_{TM}$. This is caused by the transmission loss of wave $S_t$ increasing with the thickness of slab and finally $S_{r2}$ does not exist any longer as shown in Fig.~3. From Fig.~4, we can conclude that as frequency increasing, the decay rate of reflection coefficient increases, that is, $r'_{TE}$ and $r'_{TM}$ converge to $r_{TE}$ and $r_{TM}$ faster at higher frequencies. Moreover, $r'_{TE}$ and $r'_{TM}$ at large incident angle decays faster than that at small incident angle at a certain frequency.

Based on above analysis, a new concept ``settling thickness'', which describe the minimum thickness of slab inducing reflection coefficient within a certain band, is proposed. For example, a sensing system requires variation of reflection coefficient less than 0.2~dB to achieve steady reflected wave. Fig.~5 illustrates how the settling thickness of glass at 1~THz is obtained. As shown in Fig.~5, when the slab made of glass is thicker than 1.4~mm, the fluctuating reflection coefficient curve reaches and steadies within the range from -7.54 to -7.14~dB (-7.34$\pm$0.2~dB). Therefore, any surfaces made of glass thicker than 1.4~mm, no matter what they are, e.g., window glass, mirror, empty bottle, bottle containing any liquid, will induce reflection coefficient within the given tolerance 0.2~dB at 1~THz. In general, common surfaces made of glass in indoor environment are thicker than 1.4~mm, hence induce steady reflected waves at 1~THz. However, at lower frequencies, reflecting surfaces with various thicknesses can result in considerable variations of reflection coefficient. Therefore, we can conclude that any sensing solutions based on signal strength of reflected wave are feasible at higher frequencies only. This is an important principle relating to design of RL-based sensing applications, since the thicknesses of the reflecting surfaces in an environment are entirely unpredictable. The settling thicknesses of wood, plaster, and glass at 28~GHz, 100~GHz, and 1~THz inducing variation of reflection coefficient within 0.2~dB are tabulated in TABLE~III.

\begin{figure}[htbp]
	\centerline{\includegraphics[width=\linewidth, height=10cm, keepaspectratio]{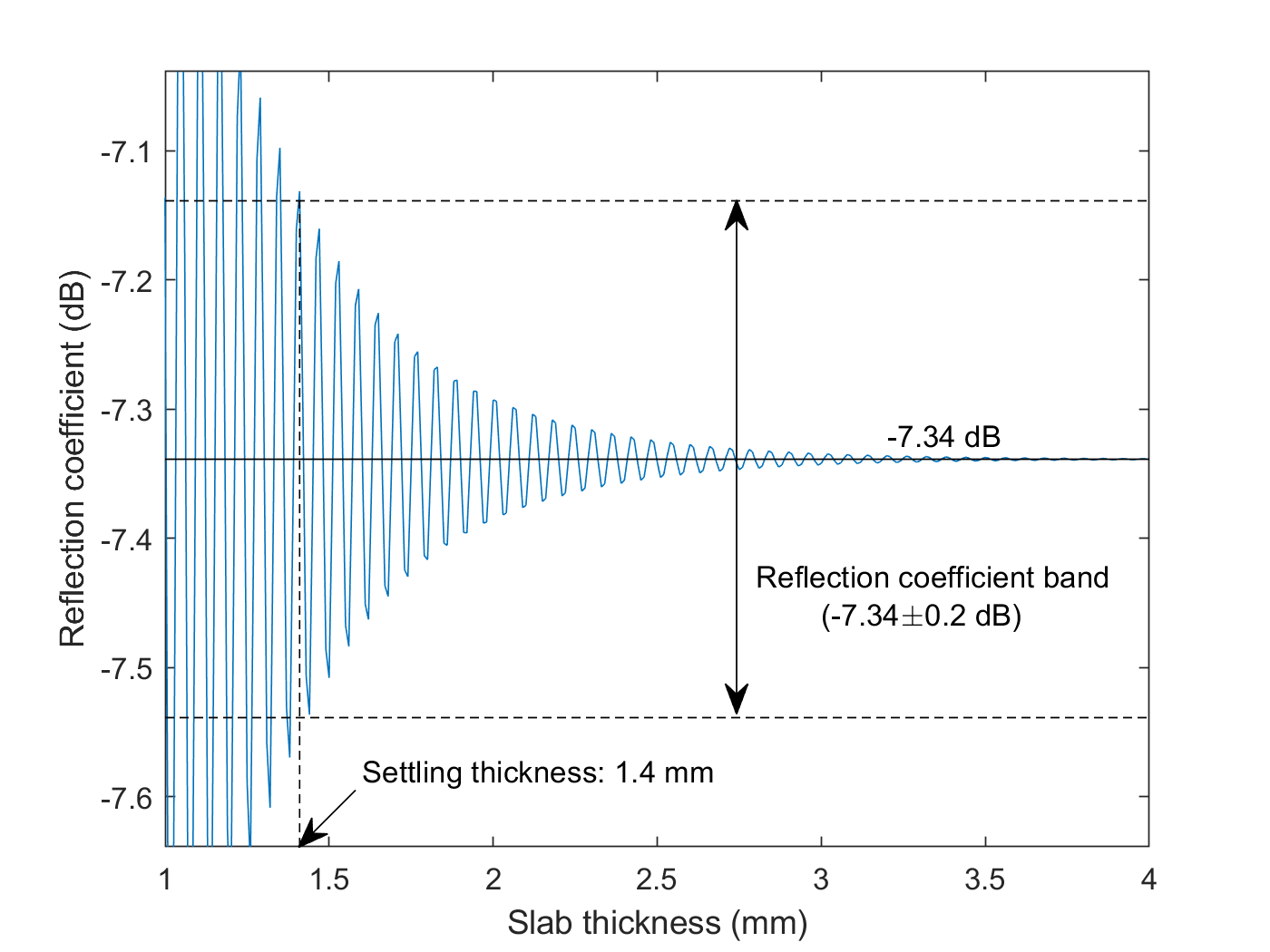}}
	\caption{An illustration of settling thickness of slab made of glass at 1~THz with incident angle 0\textdegree.}
	\label{fig_5}
\end{figure}

\begin{table}[htbp]
	\caption{Settling thicknesses of wood, plaster, and glass at 28~GHz, 100~GHz, and 1~THz inducing variation of reflection coefficient less than 0.2~dB}
	\begin{center}
		\begin{tabular}{ccccc}
			\hline
			\multirow{2}{*}{\textbf{Material}} &\multicolumn{3}{c}{\textbf{Settling thickness (mm)}} \\
			& 28~GHz & 100~GHz & 1~THz\\
			\hline
			Wood &	83 & 21 & 1.8 \\
			Plaster & 138 & 55 & 10\\
			Glass & 103 & 22 & 1.4\\
			\hline
		\end{tabular}
	\end{center}
	\label{tab_3}
\end{table} 

\section{Map-Assisted Material Identification Method}

\begin{figure*}[t]
	\centerline{\includegraphics[width=\linewidth, height=10cm, keepaspectratio]{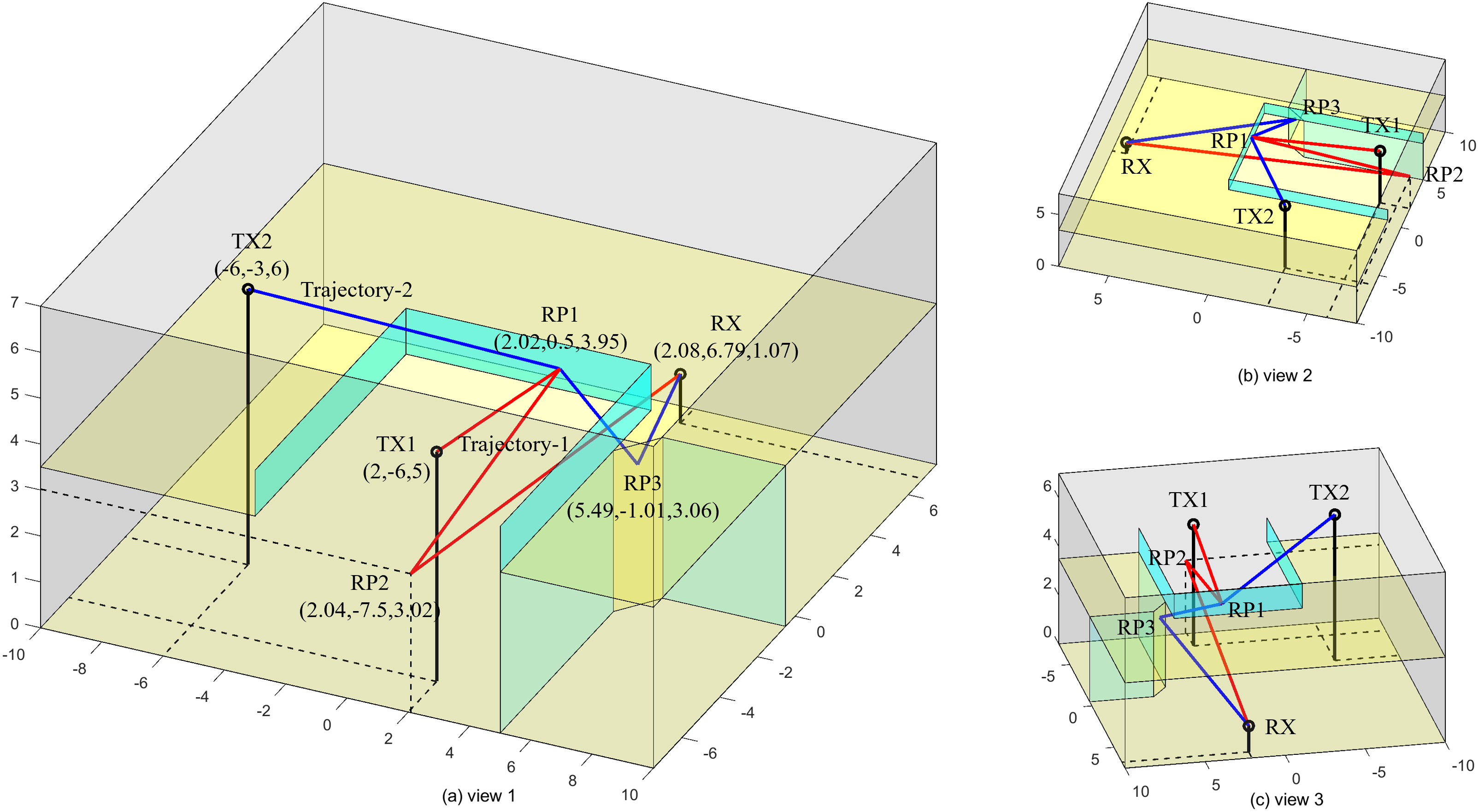}}
	\caption{Two trajectories between TX1-RX and TX2-RX in a simulated two-storey building.}
	\label{fig_6}
\end{figure*}

\begin{figure*}[t]
	\centerline{\includegraphics[width=\linewidth, height=10cm, keepaspectratio]{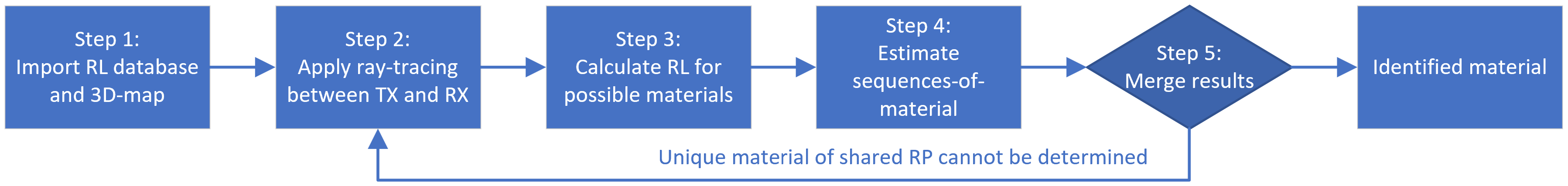}}
	\caption{Block diagram for the proposed material identification method.}
	\label{fig_7}
\end{figure*}

\begin{table*}[h]
	\caption{All possible sequences-of-material and corresponding total RL of trajectory-1 and trajectory-2}
	\begin{center}
		\begin{tabular}{cccccccccc}
			\hline
			Trajectory & first RP & second RP & {$\theta_{i1}$} & $\theta_{i2}$ & Sequence-of-material & RL1 (dB) & RL2 (dB) & Total RL (dB) \\
			\hline
			\multirow{9}{*}{Trajectory-1} & \multirow{9}{*}{RP1(2.02, 0.5, 3.95)} & \multirow{9}{*}{RP2(2.04, -7.5, 3.02)} & \multirow{9}{*}{7.9\textdegree} & \multirow{9}{*}{7.2\textdegree} & (RP1-wood, RP2-wood) & 16.39 & 16.4 & 32.79 \\
			& & & & & (RP1-plaster, RP2-plaster) & 11.86 & 11.86 & 23.72 \\
			& & & & & (RP1-glass, RP2-glass) & 7.34 & 7.34 & 14.68 \\
			& & & & & (RP1-wood, RP2-plaster) & 16.39 & 11.86 & 28.25 \\
			& & & & & (RP1-plaster, RP2-wood) & 11.86 & 16.4 & 28.26 \\
			& & & & & (RP1-wood, RP2-glass) & 16.39 & 7.34 & 23.73 \\
			& & & & & (RP1-glass, RP2-wood) & 7.34 & 16.4 & 23.74 \\
			& & & & & \textbf{(RP1-plaster, RP2-glass)}$^a$ & \textbf{11.86} & \textbf{7.34} & \textbf{19.2} \\
			& & & & & \textbf{(RP1-glass, RP2-plaster)}$^a$ & \textbf{7.34} & \textbf{11.86} & \textbf{19.2} \\
			\hline
			\multirow{9}{*}{Trajectory-2} & \multirow{9}{*}{RP1(2.02, 0.5, 3.95)} & \multirow{9}{*}{RP3(5.49, -1.01, 3.06)}& \multirow{9}{*}{67.1\textdegree} & \multirow{9}{*}{25\textdegree} & (RP1-wood, RP3-wood) & 9.3 & 16.13 & 25.43 \\
			& & & & & (RP1-plaster, RP3-plaster) & 7.67 & 11.76 & 19.43 \\
			& & & & & (RP1-glass, RP3-glass) & 5.93 & 7.31 & 13.24 \\
			& & & & & \textbf{(RP1-wood, RP3-plaster)}$^b$ & \textbf{9.3} & \textbf{11.76} & \textbf{21.06} \\
			& & & & & (RP1-plaster, RP3-wood) & 7.67 & 16.13 & 23.8 \\
			& & & & & (RP1-wood, RP3-glass) & 9.3 & 7.31 & 16.61 \\
			& & & & & \textbf{(RP1-glass, RP3-wood)}$^b$ & \textbf{5.93} & \textbf{16.13} & \textbf{22.06} \\
			& & & & & (RP1-plaster, RP3-glass) & 7.67 & 7.31 & 14.98 \\	
			& & & & & (RP1-glass, RP3-plaster) & 5.93 & 11.76 & 17.69 \\
			\hline
		\end{tabular}
	\end{center}
	\label{tab_4}
	\footnotesize{$^a$ Two sequences-of-material of trajectory-1 inducing total RL within 19$\pm$1~dB}\\
	\footnotesize{$^b$ Two sequences-of-material of trajectory-2 inducing total RL within 21.5$\pm$1~dB}\\
\end{table*}

As mentioned in Section~II, the prior methods of material identification suffer from some serious disadvantages. To tackle them, a novel method for identifying materials of a number of reflecting surfaces along multiple-bounce-reflection trajectories is proposed. The method is a mixed methodology based on the method proposed in \cite{b4} and ray-tracing technology. A simulated two-storey building depicted in Fig.~6 is used to present the method. Sub-figures \mbox{(a)-(c)} in Fig.~6 illustrate the same building from different views, the aim of the sub-figures is to show the spatial relations of the 3D objects in the building. The dimensions of the building are 20~m$\times$15~m$\times$7~m. There is a rectangular cubicle with a beveled door on the first floor. The second floor has an aperture with size of \mbox{8$\times$8~m}, the aperture is blocked by the vertical railings with 1~m height. The purpose of the aperture and the railings is to provide rich scattering condition between TX antennas mounted on the ceiling and movable RX devices on the first floor. The floor and the beveled door in yellow are made of wood, the building walls in gray are made of plaster, the railings and cubicle walls in cyan are made of glass. The objects in the building are thicker than corresponding settling thicknesses, therefore, all reflecting surfaces induce steady reflection coefficient no matter how thick they are. The step-by-step procedure of proposed method (see block diagram in Fig.~7) is as follows:

Step~1: By a large amount of empirical measurements or theoretical calculations, RLs induced by reflecting surfaces made of different materials at various frequencies and at any incident angles are obtained and stored in a RL database. TABLE~I gives an example of RL database. Moreover, the 3D-map of the environment is imported. The 3D-map should at least contain the 3D geometric information for each major structure of buildings or rooms and for any objects in the scenario.

Step~2: Apply ray-tracing to each TX-RX association. The output of step~2 is a set of trajectories between TX and RX and trajectory information including coordinate of each RP and incident angle of each reflection. For example, in Fig.~6, two double-bounce-reflection trajectories \mbox{TX1$\rightarrow$RP1$\rightarrow$RP2$\rightarrow$RX} and \mbox{TX2$\rightarrow$RP1$\rightarrow$RP3$\rightarrow$RX} (referred to as ``trajectory-1'' and ``trajectory-2'' below) are identified. By geometric calculations, the coordinate of each RP as well as incident angle of each reflection of trajectory-1 and 2 are obtained and tabulated in TABLE~IV.

Step~3: Calculate RL of each reflection on the trajectories obtained in Step~2 for all possible materials, then list the total RL induced by reflecting surfaces along multiple-bounce-reflection trajectory for all possible sequences-of-material. A sequence-of-material, which contains an ordering of a few materials, represents the material of reflecting surfaces in temporal order when the radio wave strikes the surfaces in sequence. If m1 and m2 are two materials, we use \mbox{(RPx-m1, RPy-m2)} to denote the sequence-of-material of a double-bounce-reflection in which the first reflecting surface at RPx is made of m1, and the second reflecting surface at RPy is made of m2. For simplicity we assume that the objects in Fig.~6 can be made of one of wood, plaster, and glass only. Therefore, there are nine possible sequences-of-material for two reflecting surfaces in total for double-bounce-reflection trajectories. By performing RL calculation method proposed in \cite{b4} at 100~GHz, for all nine possible sequences-of-material, RL induced by the first reflecting surface (RL1) at incident angle $\theta_{i1}$, RL induced by the second reflecting surface (RL2) at incident angle $\theta_{i2}$, and the total RL (RL1 plus RL2) of both trajectories are summarized in TABLE~IV.

Step~4: Transmit radio signals along the trajectories obtained in Step~2 and measure the total RL induced by multiple reflecting surfaces. By looking up the total RL in the table created in Step~3, the sequence-of-material can be determined or narrowed down. For example, in Fig.~6, if the measured total RL of trajectory-1 is 19~dB, and the maximum measurement uncertainty is 1~dB, that is, the true total RL of trajectory-1 ranges from 18 to 20~dB (19$\pm$1~dB). By looking up TABLE~IV, two sequences-of-material inducing total RL within 19$\pm$1~dB are found: (RP1-plaster, RP2-glass) and (RP1-glass, RP2-plaster) with footnote $^a$ in TABLE~IV. Similarly, the measured total RL of trajectory-2 is 21.5~dB, by looking up total RL within 21.5$\pm$1~dB in TABLE~IV, the possible sequences-of-material of trajectory-2 could be (RP1-wood, RP3-plaster) and (RP1-glass, RP3-wood) with footnote $^b$ in TABLE~IV. These four sequences-of-material are listed in TABLE~V. Compared to the sequences-of-material in TABLE~IV, the possible sequences-of-material of trajectories-1 and 2 have narrowed significantly but we still cannot conclude the materials of reflecting surfaces at RP1, RP2, and RP3.

Step~5: Check the RP that is shared among multiple trajectories obtained in Step~2, then merge the possible sequences-of-material obtained by Step~4 to determine or narrow down the material of common RP shared by multiple trajectories. If there is no RP shared among the trajectories, or unique material of shared RP cannot be determined by Step~5, go to Step~2 and execute Step~2-5 repeatedly for TX/RX at other positions, to create additional trajectories that have shared RPs. For example, trajectory-1 and trajectory-2 share a common reflecting surface at RP1\mbox{(2.02,0.5,3.95)}. From the possible sequences-of-material listed in TABLE~V, glass is identified as the material of reflecting surface at RP1 not only by total RL of trajectory-1, but also by total RL of trajectory-2. Since trajectory-1 and 2 are independent, we can unambiguously conclude that reflecting surface at RP1 is made of glass, and we are consequently able to conclude that reflecting surfaces at RP2 and RP3 are made of plaster and wood, respectively.

\begin{table}[htbp]
	\caption{The possible materials of RP1-RP3 obtained by two independent RL measurements for trajectory-1 and 2}
	\begin{center}
		\begin{tabular}{ccccc}
			\hline
			\multirow{2}{*}{\textbf{Trajectory}} &\multicolumn{3}{c}{\textbf{Possible materials of RP1-RP3}} \\
			& RP1 & RP2 & RP3\\
			\hline
			Trajectory-1 &	plaster & glass & \\
			Trajectory-1 & \textbf{glass} & plaster & \\
			\hline
			Trajectory-2 & wood &  & plaster\\
			Trajectory-2 & \textbf{glass} &  & wood\\
			\hline
		\end{tabular}
	\end{center}
	\label{tab_5}
\end{table} 

\section{Conclusions}
In this paper, a new concept ``settling thickness'' is proposed, and the settling thicknesses of three materials are calculated at 28~GHz, 100~GHz, and 1~THz. From the calculation results, it is apparent that for 100~GHz and above the common reflecting surfaces in indoor environment have enough settling thickness to induce steady reflection coefficient. We then propose a novel map-assisted material identification method. Simulation results show that the proposed method can determine the materials of reflecting surfaces along multiple-bounce-reflection trajectories with a certain measurement uncertainty.

\vspace{12pt}
\color{red}
\end{document}